\documentclass{article}
\usepackage[T1]{fontenc} 
\usepackage[utf8]{inputenc} 
\usepackage{ismir,amsmath,cite,url}
\usepackage{graphicx}
\usepackage{color}
\usepackage{booktabs}
\usepackage[table,xcdraw]{xcolor}
\usepackage{multirow}

\title{Transfer Learning of wav2vec 2.0 for Automatic Lyric Transcription}

\oneauthor
{Longshen Ou$^*$, Xiangming Gu$^*$, Ye Wang}
{School of Computing, National University of Singapore \\ 
\texttt{\{longshen, xiangming, wangye\}@comp.nus.edu.sg} \\
$^*$Both authors contributed equally to this research.} 

\def\authorname{Longshen Ou, Xiangming Gu, Ye Wang}

\begin{document}

\maketitle

\begin{abstract}

Automatic speech recognition (ASR) has progressed significantly in recent years due to the emergence of large-scale datasets and the self-supervised learning (SSL) paradigm. However, as its counterpart problem in the singing domain, the development of automatic lyric transcription (ALT) suffers from limited data and degraded intelligibility of sung lyrics. To fill in the performance gap between ALT and ASR, we attempt to exploit the similarities between speech and singing. In this work, we propose a transfer-learning-based ALT solution that takes advantage of these similarities by adapting wav2vec 2.0, an SSL ASR model, to the singing domain. We maximize the effectiveness of transfer learning by exploring the influence of different transfer starting points. We further enhance the performance by extending the original CTC model to a hybrid CTC/attention model. Our method surpasses previous approaches by a large margin on various ALT benchmark datasets. Further experiments show that, with even a tiny proportion of training data, our method still achieves competitive performance.

\end{abstract}

\section{Introduction}
\label{sec:introduction}

Automatic lyric transcription (ALT) systems allow for lyrics to be obtained from large musical datasets without requiring laborious manual transcription. These lyrics can then be used for many music information retrieval (MIR) tasks, including query by singing \cite{hosoya2005querybysing}, audio indexing \cite{fujihara2008audioindexing}, etc. Besides, because lyric alignment systems are typically built upon ALT models \cite{gupta2020automatic, demirel2021low}, a strong-performing ALT model can lay a solid foundation for better audio-to-text alignment performance. Consequently, ALT is becoming an increasingly active topic in the recent MIR community.

One option for improving ALT performance is to incorporate knowledge obtained from studies involving the transcription of speech. Indeed, ALT is usually treated as a separate problem from automatic speech recognition (ASR), e.g., \cite{dabike2019data:dsing, demirel2020automatic, gupta2020automatic, demirel2021mstre, gao2022genre}, eschewing large-scale speech datasets and well-developed ASR systems. However, the absence of large-scale singing datasets has impeded the construction of high-performing ALT models. While there are distinctions between sung and spoken language, e.g., sung language being less intelligible and hence harder to recognize \cite{dabike2019data:dsing, sharma2019sing_eval}, they share many similarities, such as having the same vocabularies and being produced by similar physical mechanisms. Therefore, we believe it is worth investigating whether we can use knowledge and datasets from the speech domain to compensate for the inadequacy of singing datasets and bolster the performance of ALT systems.

Transfer learning methods have been found to effectively alleviate the requirement for a large amount of training data for some low-resource tasks \cite{torrey2009transfer, pan2009survey}. For example, speech recognition for non-native speakers \cite{sullivan2022improving}, and machine translation for low-resource languages \cite{zoph2016transfer}. In such scenarios, transfer learning helps mitigate the problem of insufficient data by adapting data and knowledge from related high-resource tasks or domains. 

In recent years, self-supervised learning (SSL) has become a new paradigm in ASR research. Several SSL methods can perform excellently with access to only a few hours or even a few minutes of labeled data \cite{hsu2021hubert, schneider2019wav2vec, baevski2020wav2vec}. Among them, wav2vec 2.0 \cite{baevski2020wav2vec} has been shown to be a particularly promising model for transfer learning \cite{sullivan2022improving}. wav2vec 2.0 is an effective few-shot learner that only requires a small amount of data from the target domain or problem to achieve impressive results \cite{riviere2020unsupervised, khurana2022magic}. This property makes wav2vec 2.0 a promising candidate to help ALT systems overcome the issue of limited training data by transferring speech representation knowledge to the singing domain.

The contributions of this paper contain four aspects:

\begin{itemize}

\item We propose an ALT solution that takes advantage of the similarities between spoken and singing voices. This is achieved by performing transfer learning using wav2vec 2.0 on singing data after pretraining and finetuning on speech data.

\item We maximize the effectiveness of transfer learning by exploring the influence of different transfer starting points. We show that both pretraining and finetuning on speech data contribute to the high performance of our ALT system. 

\item We further enhance the system's performance by extending the original connectionist temporal classification (CTC) model to a hybrid CTC/attention model for better convergence and more accurate decoding.

\item Our method surpasses previous ones on various benchmark ALT datasets, including DSing \cite{dabike2019data:dsing}, DALI \cite{meseguer2019data:dali1, meseguer2020data:dali2}, Jamendo \cite{stoller2019end}, Hansen \cite{hansen2012recognition}, and Mauch \cite{mauch2011integrating}, by about 25\% relative WER reduction on average. We further show that with less than one-tenth of labeled singing data, our method can still achieve state-of-the-art results on the test split of DSing, demonstrating its effectiveness in low-resource ALT setups.

\end{itemize}

\begin{figure*}[t!]
\begin{center}
\includegraphics[width=0.95\linewidth]{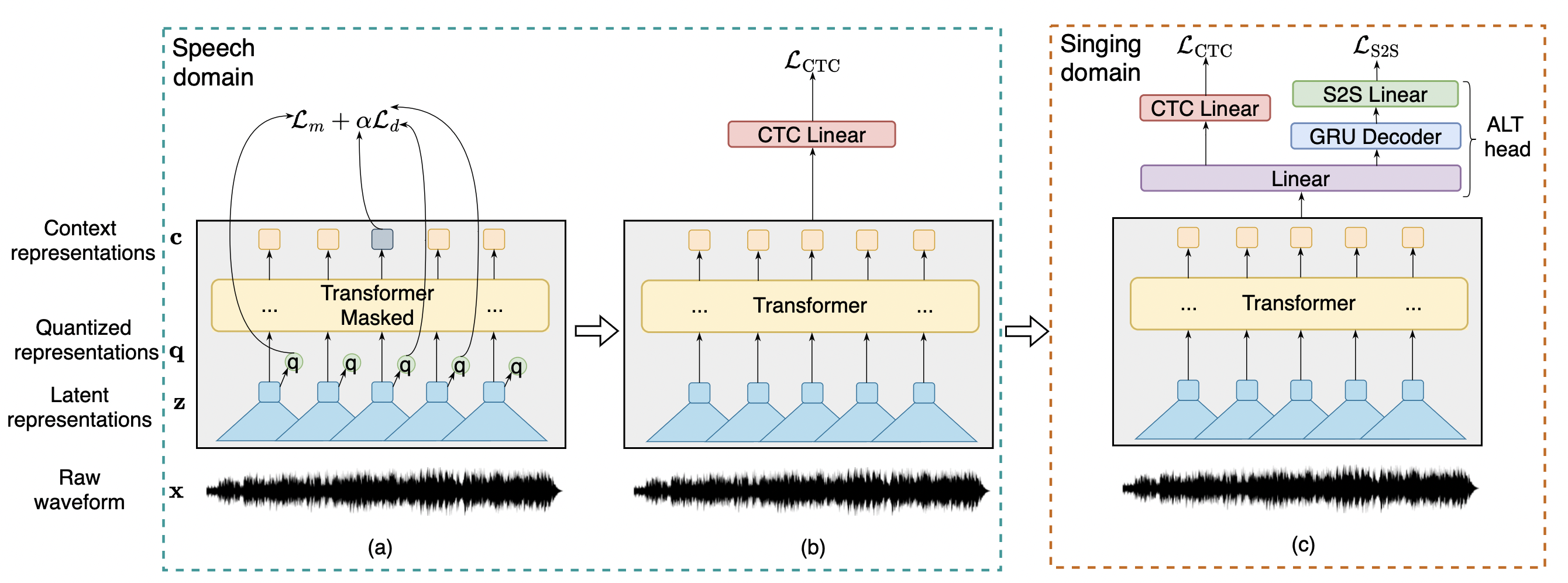}
\caption{An overview of our training framework. (a) Stage I: Pretraining on speech data. (b) Stage II: Finetuning on speech data. (c) Stage III: Transferring on singing data.}
\label{fig1}
\end{center}
\vskip -0.1in
\end{figure*}
\section{Related Work}
\label{sec:related_works}

\subsection{Automatic Lyric Transcription}
Recent progress in lyric transcription has been mainly driven by three factors. First, the construction and curation of datasets containing aligned audio and lyrics, including DAMP Sing! 300x30x2 \cite{data:smule, dabike2019data:dsing} and DALI \cite{meseguer2019data:dali1, meseguer2020data:dali2}, lay the foundation for data-driven ALT models. Second, the design of ALT acoustic models benefits from architectures of automatic speech recognition (ASR) models and can be further improved by adopting singing domain knowledge as inductive bias. Representative work includes TDNN-F with its variants \cite{dabike2019data:dsing, demirel2020automatic, gupta2020automatic, demirel2021mstre} and vanilla/convolution-augmented Transformers \cite{basak2021end, zhang2021pdaugment, gao2022genre}. Additionally, \cite{gu2022mm} proposed to leverage the complementary information of additional modalities (video and wearable IMU sensors) for ALT systems. Third, through data augmentation methods such as adjusting speech data to make it more ``song-like'' \cite{kruspe2015songified} or synthesizing singing voice from speech voice \cite{basak2021end, zhang2021pdaugment}, more training data can be created for ALT models, thus alleviating the data sparsity problem. 

\subsection{Self-supervised speech representation learning}
The success of deep learning methods is highly related to the power of the learned representations. Although supervised learning still dominates the speech representation learning field, it has several drawbacks. For example, substantial amounts of labeled data are required to train supervised learning ASR models \cite{schneider2019wav2vec, baevski2020wav2vec}. Moreover, representations obtained through supervised learning tend to be biased to specific problems, thus are difficult to extend to other applications \cite{pascual2019learning}.

To mitigate the above problems, a series of self-supervised learning (SSL) frameworks for speech representation learning have emerged, e.g., Autoregressive Predictive Coding (APC) \cite{chung2019unsupervised}, Contrastive Predictive Coding (CPC) \cite{schneider2019wav2vec}, and Masked Predictive Coding (MPC) \cite{jiang2019improving, zhang2021transformer, hsu2021hubert}. Moreover, wav2vec 2.0 takes advantage of both CPC and MPC to conduct self-supervised learning and has become the new paradigm for the ASR task \cite{baevski2020wav2vec}. wav2vec 2.0 has been also widely adopted as the feature extractor for other speech-related applications, e.g. speech emotion recognition \cite{pepino2021emotion,zhao2022memobert}, keyword spotting \cite{seo2021wav2kws}, speaker verification and language identification \cite{fan2020exploring}, demonstrating that speech representations learned from wav2vec 2.0 are robust and transferable for downstream tasks.

\section{Methodology}\label{sec:methodology}
In this section, we firstly recap the structure of wav2vec 2.0 \cite{baevski2020wav2vec}. Then we elaborate on the three training stages of the proposed methods, including pretraining on speech data, finetuning on speech data, and transferring to the singing domain. 

\subsection{Structure of wav2vec 2.0}
As shown in Fig. \ref{fig1}, wav2vec 2.0 is built with a CNN-based feature encoder, a Transformer-based context network, and a quantization module. For raw audio inputs $\bm{x}$ with a sampling rate of 16 kHz, the feature encoder accepts $\bm{x}$ and obtains the latent speech representations $\bm{z}$. The feature encoder has seven blocks, each of which includes a 1D temporal convolution with 512 channels followed by layer normalization \cite{ba2016layer} and GELU activation \cite{hendrycks2016gaussian}. Consequently, $\bm{z}\in \mathcal{R}^{T\times1024}$ are 2D representations with a frequency of 49 Hz. To exploit the temporal relationship among different frames of latent representations, $\bm{z}$ are further fed into the context network, which is parameterized by 12 Transformer blocks \cite{vaswani2017attention}. Each block has a multi-head attention module with 16 attention heads and a Feed-Forward Network (FFN) with 4,096 hidden dimensions. Resulting context representations $\bm{c}\in\mathcal{R}^{T\times1024}$ are the features extracted from the audio signal and used for downstream tasks.

In addition to being fed into the context network, $\bm{z}$ are also accepted by a quantization module, which learns quantized speech representations $\bm{q}$, thus facilitating the self-supervised training. 


\subsection{Stage I: Pretraining on speech data}
wav2vec 2.0 is pretrained through an SSL method \cite{baevski2020wav2vec} on large-scale unlabeled speech data, as displayed in Fig. \ref{fig1}(a). Before latent representations $\bm{z}$ are fed into the context network, several consecutive frame sequences are randomly masked. The masked frames are replaced by a trainable vector. wav2vec 2.0 is trained by optimizing the combination of contrastive loss and diversity loss $\mathcal{L}_m+\alpha\mathcal{L}_d$ ($\alpha$ refers to a balancing hyper-parameter). The contrastive objective is defined as:
\begin{equation}
    \mathcal{L}_m=-\log\frac{\exp(sim(\bm{c}_t, \bm{q}_t)/\kappa)}{\sum_{\bm{\widetilde{q}}\sim \bm{Q}_t}\exp(sim(\bm{c}_t, \bm{\widetilde{q}}_t)/\kappa)}
\end{equation}
where $\bm{c}_t$ is $t$-th frame of context representations, $\bm{Q}_t$ represents all possible quantized representations, the temperature value $\kappa$ is set as $0.1$ and $sim$ refers to the cosine similarity. The diversity loss $\mathcal{L}_d$ is designed to encourage the usage of all entries in codebooks. We refer readers to \cite{baevski2020wav2vec} for more details.

\subsection{Stage II: Finetuning on speech data}
The process of finetuning requires labeled speech data. As shown in Fig. \ref{fig1}(b), the quantization module in wav2vec 2.0 is disabled since it is only used in Stage I, and a linear layer (CTC linear) is added on top of the Transformer. The whole model is trained by optimizing the connectionist temporal classification (CTC) loss $\mathcal{L}_{\text{CTC}}$ \cite{graves2006connectionist}. Suppose the ground-truth transcription is $\bm{w}^*$, which is a sequence of character tokens. The CTC loss is defined as:
\begin{equation}
    \mathcal{L}_{\text{CTC}}=-\log\sum_{\bm{\pi}\in\mathcal{B}^{-1}(\bm{w}^*)}\prod_{t=1}^Tp(\bm{\pi}_t|\bm{f}_t)
\end{equation}
where $\bm{f}_t$ refers to $\bm{c}_t$ in this stage, $T$ is the number of frames, $\mathcal{B}$ is a function to map an alignment sequence $\bm{\pi}_{1:T}$ to $\bm{w}_{1:N}^*$ (where $N$ represents the number of character tokens) by removing duplicate characters and blanks while its inverse function $\mathcal{B}^{-1}(\bm{w}^*)$ refers to all the CTC paths mapped from $\bm{w}^*$. For speech data, there are 31 tokens for character targets, including 26 letters, the quotation mark, a word boundary token, $<bos>$, $<eos>$, and CTC blank token. The probability $p(\bm{\pi}_t|\bm{f}_t)$ is computed by the CTC linear layer followed by a softmax operation. Besides the supervised CTC loss, pseudo-labeling is also adopted during finetuning. Please refer to \cite{baevski2020wav2vec, xu2021self} for more details. 


\subsection{Stage III: Transferring on singing data}
\subsubsection{From CTC to CTC/attention}
To transfer the trained wav2vec 2.0 from the speech domain to the singing domain, we retain the weights of the feature encoder and the context network after Stages I and II. Furthermore, inspired by \cite{watanabe2017hybrid}, we extend the original CTC system into a hybrid CTC/attention system through the addition of an ALT head on top of wav2vec 2.0 instead of a single CTC linear layer (Fig. \ref{fig1}(c)). 

The context representations $\bm{c}$ are first fed into a linear layer followed by a leaky ReLU activation layer to obtain the features $\bm{f}$. Then $\bm{f}$ are sent to two network branches. One branch is a CTC linear layer, which aims to compute $p(\bm{\pi}_t|\bm{f}_t)$ as explained in sec. 3.2.2. Another branch is an attention-based GRU decoder \cite{chorowski2015attention} followed by a sequence-to-sequence (S2S) linear layer. The GRU decoder has a single layer with a hidden dimension of 1,024 and utilizes location-aware attention \cite{chorowski2015attention} with attention dimension 256. The decoder and S2S linear layer autoregressively compute the probability $p(\bm{w}_n|\bm{w}_{<n}, \bm{f}_{1:T}),\, n=1,2,...,N$.

\subsubsection{Training and Evaluation}

During Stage III, wav2vec 2.0 and the ALT head are trained through the combination of CTC loss \cite{graves2006connectionist} and S2S loss \cite{bahdanau2016end}:
\begin{equation}\label{eq3}
    \mathcal{L}_w=\lambda_a \mathcal{L}_{\text{CTC}}+ (1-\lambda_a)\mathcal{L}_{\text{S2S}}
\end{equation}
\begin{equation}
    \mathcal{L}_{\text{S2S}} =  -\log \prod_{n=1}^N p(\bm{w}_n^*|\bm{w}_{<n}^*, \bm{f}_{1:T})
\end{equation}
where $\lambda_a$ is a hyper-parameter to balance the CTC loss term and S2S loss term. To overcome catastrophic forgetting, we adopt a smaller learning rate for wav2vec 2.0 compared to the ALT head. 

To evaluate the performance of the trained model, the most likely lyrics are predicted using beam search:
\begin{align}\label{eq5} \bm{w}'&=\arg\max_{\bm{w}}\lambda_b\log\sum_{\bm{\pi}\in\mathcal{B}^{-1}(\bm{w})}\prod_{t=1}^Tp(\bm{\pi}_t|\bm{f}_t)\nonumber\\
    &+(1-\lambda_b)\log \prod_{s=1}^S p(\bm{w}_s|\bm{w}_{<s}, \bm{f}_{1:T})\nonumber\\
    &+\lambda_c\log p_{LM}(\bm{w})
\end{align}
where $\lambda_b$ and $\lambda_c$ are two hyper-parameters in the decoding process. The language model is implemented by a 3-layer LSTM. The characters are firstly projected to embeddings and then fed into the LSTM with a hidden dimension of 2,048 to obtain RNN features. Finally, the RNN features are accepted by a 3-layer MLP with a hidden dimension of 1,024 to output the probability $p_{LM}(\bm{w})$. 

\section{Experiments}
\label{sec:experiments}

\subsection{Datasets and preprocessing}
\label{sec:dataset}

We use various accessible mainstream lyric transcription datasets for our experiments, including DALI \cite{meseguer2019data:dali1, meseguer2020data:dali2}, Hansen \cite{hansen2012recognition}, Mauch \cite{mauch2011integrating},  Jamendo \cite{stoller2019end}, and a curated version of DAMP Sing! 300x30x2 \cite{data:smule} called DSing \cite{dabike2019data:dsing}. The train/development/test splits in our experiments are defined as follows. For the DSing dataset, we use the same split configuration as in \cite{dabike2019data:dsing}. Specifically, there are three different sizes of training sets (DSing1, DSing3, DSing30), as well as a development set DSing\textsuperscript{dev} and a test set DSing\textsuperscript{test}. As for the DALI dataset, we divide all publicly available audio in DALI v2 \cite{meseguer2020data:dali2} into training and development subsets (DALI\textsuperscript{train} and DALI\textsuperscript{dev} respectively).\footnote{At the time of this research, some audios are not retrievable through their YouTube links in the public-available metadata. Although audios containing the same titles and artist names can be found online, we cannot guarantee they perfectly match the annotations for the original audio versions in DALI. Discarding invalid audio is only performed for the training and the development sets.} We use DALI\textsuperscript{test} \cite{demirel2021mstre}, a subset of DALI v1 \cite{meseguer2019data:dali1}, as a test set for experiments involving DALI. The full Hansen, Mauch, and Jamendo datasets are used as additional out-of-domain test sets. In the training and development splits of the DALI dataset, songs that overlapped with any of our test sets are removed for more objective testing.


\begin{table}[t!]
\centering
\begin{tabular}{l|l|cc}
\hline
 Split & Dataset & \# Utt. & Total Dur. \\ \hline
 & DSing1 & 8,794 & 15.1 h \\
\multicolumn{1}{c|}{\multirow{2}{*}{Train}} & DSing3 & 25,526 & 44.7 h \\
\multicolumn{1}{c|}{} & DSing30 & 81,092 & 149.1 h \\
 & DALI\textsuperscript{train} & 268,392 & 183.8 h \\ \hline
\multicolumn{1}{c|}{\multirow{2}{*}{Dev}} & DSing\textsuperscript{dev} & 482 & 41 min \\
\multicolumn{1}{c|}{} & DALI\textsuperscript{dev} & 1,313 & 55 min \\ \hline
 & DSing\textsuperscript{test} & 480 & 48 min \\
 & DALI\textsuperscript{test} & 12,471 & 9 h \\
\multicolumn{1}{c|}{Test} & Jamendo & 921 & 49 min \\
 & Hansen & 634 & 34 min \\
 & Mauch & 878 & 54 min \\ \hline
\end{tabular}
\caption{Statistics of segmented utterance-level datasets. }
\label{tab:data}
\end{table} 

The speech data used to pretrain and finetune the wav2vec 2.0 is initially monophonic. Therefore, we expect wav2vec 2.0 to extract better representation features for singing data if monophonic audios are given as the inputs. Thus, we extract vocal parts from all of the polyphonic recordings in DALI, Mauch, and Jamendo datasets using Demucs v3 \textit{mdx\_extra}\cite{defossez2019demucs}, which is the state-of-the-art source separation model that achieved the first rank at the 2021 Sony Music DemiXing Challenge (MDX). This ensures that we are consistent with the input requirements of wav2vec 2.0 and minimize the interference of musical accompaniment. We adopt utterance-level input for both training and testing. To facilitate the experiments, we perform utterance-level segmentation on all audios according to their annotations. Utterances with obvious faulty annotations are removed (e.g., utterances labeled as several words but have shorter than 0.1 s duration). The statistics of all datasets after utterance-level segmentation are listed in Table \ref{tab:data}. The ``total duration'' column refers to the sum of durations of all utterances in the datasets, excluding the instrumental-only parts between utterances, hence resulting in shorter durations than \cite{demirel2021mstre}. We notice that the average utterance duration of DSing is longer than DALI (6.52 s vs. 2.47 s). Finally, lyric texts in training sets are normalized by converting all letters to upper case, converting digits to words, discarding out-of-vocabulary characters, discarding meaningless lines (e.g., ``**guitar solo**''), and removing redundant space. 

\begin{table*}[t!]
\centering
\begin{tabular}{@{}l|cccccc@{}}
\toprule
Method & \multicolumn{1}{l}{DSing\textsuperscript{dev}} & \multicolumn{1}{l}{DSing\textsuperscript{test}} & \multicolumn{1}{l}{DALI\textsuperscript{test}} & \multicolumn{1}{l}{Jamendo} & \multicolumn{1}{l}{Hansen} & \multicolumn{1}{l}{Mauch} \\ \midrule
TDNN-F \cite{dabike2019data:dsing} & 23.33 & 19.60 & 67.12 & 76.37 & 77.59 & 76.98 \\
CTDNN-SA \cite{demirel2020automatic} & \underline{17.70} & \underline{14.96} & 76.72 & 66.96 & 78.53 & 78.50 \\
Genre-informed AM \cite{gupta2020automatic} & - & 56.90 & - & 50.64 & 39.00 & 40.43 \\
MSTRE-Net \cite{demirel2021mstre} & - & 15.38 & \underline{42.11} & \underline{34.94} & \underline{36.78} & \underline{37.33} \\
DE2 - segmented \cite{demirel2021low} & - & - & - & 44.52 & 49.92 & - \\ \midrule
Ours & \textbf{12.34} & \textbf{12.99} & \textbf{30.85} & \textbf{33.13} & \textbf{18.71} & \textbf{28.48} \\ \bottomrule
\end{tabular}
\caption{WERs (\%) of various ALT systems on different singing datasets. ``-'' refers to ``non-applicable''. We use \textbf{bold face} to highlight the best results, and \underline{underline} to mark the second-best results. Note that the results of \cite{demirel2021mstre, demirel2020automatic, dabike2019data:dsing} on DALI\textsuperscript{test}, Jamendo, Hansen, and Mauch datasets are obtained without utterance segmentation.}
\label{tab:sota}
\end{table*}

\subsection{Experiment setup}
Our experiments are conducted through SpeechBrain toolkit \cite{ravanelli2021speechbrain}\footnote{Our code is released at https://github.com/guxm2021/ALT\_SpeechBrain}. Before transferring on singing data, the wav2vec 2.0 has been pretrained on LibriVox (LV-60K) and finetuned on LibriSpeech (LS-960)\footnote{https://huggingface.co/facebook/wav2vec2-large-960h-lv60-self} \cite{baevski2020wav2vec}. Then we randomly initialize the ALT head. During Stage III, we downsample all audios to 16 kHz and convert them to mono-channel by averaging the two channels of stereo audio signals. Then the singing data is augmented through SpecAugment \cite{park2019specaugment}. wav2vec 2.0 and ALT head are trained using Adam optimizer \cite{kingma2014adam}. The initial learning rates of ALT head and wav2vec 2.0 are $3\times10^{-4}$ and $1\times10^{-5}$ respectively. Learning rates are scheduled using the Newbob technique, with annealing factors of 0.8 and 0.9 respectively. The batch size is set as 4, and hyper-parameter $\lambda_a$ is set as 0.2 during the experiments. We conduct our experiments on 4 RTX A5000 GPUs. Utterances whose duration is longer than 28 seconds are filtered out during training to prevent the out-of-memory issue. This filtering is not performed during the evaluation for a fair comparison with other existing methods. 

We firstly utilize DSing30 to train the whole model for 10 epochs. We evaluate the performance of our model on DSing\textsuperscript{dev} after each epoch. Finally, the best model is selected to be evaluated on the test split DSing\textsuperscript{test}. Word error rate (WER) is adopted as the evaluation metric. During the evaluation, WERs are averaged over all utterances in a test set. The RNNLM is trained for 20 epochs using an Adam optimizer \cite{kingma2014adam} on texts of DSing30 split and validated on the DSing\textsuperscript{dev} split after each epoch. The learning rate is $1\times10^{-3}$ and the batch size is 20. During the decoding, the beam size is 512, and the hyper-parameters $\lambda_b$ and $\lambda_c$ are 0.4 and 0.5, respectively. The trained model is evaluated on the DSing\textsuperscript{test} set.

For other test splits, we adopt both DSing\textsuperscript{train} and DALI\textsuperscript{train} to train the whole model. Since these two datasets are collected from different domains, we adopt a consecutive training strategy instead of training together in order to reduce the difficulty of training. Specifically, we continue training the whole model on DALI\textsuperscript{train} split for 4 epochs. The weights of learnable parameters are initialized using the model trained on DSing30. After training, we evaluate the model on DALI\textsuperscript{test} as well as the Hansen, Mauch, and Jamendo datasets. The configuration of RNNLM is the same as above, except that we utilize texts of both DSing30 and DALI\textsuperscript{train} to train the model. To decode the lyrics, we set the beam size as 512 and the hyper-parameters $\lambda_b$ and $\lambda_c$ as 0.3 and 0.2, respectively.
\section{Results}

In this section, we firstly compare our method with state-of-the-art ALT systems on multiple benchmark datasets. Then we conduct extensive ablation studies to show the benefits of our design choices on the DSing dataset, which has more accurate manual annotations on its development and test splits. Finally, we show that our method is still effective with a limited amount of singing data.

\subsection{Comparison with the state-of-the-art}

We compare the performance of the proposed method with previous approaches, as shown in Table \ref{tab:sota}. Our method outperforms all previously published results on all the evaluation datasets. Our method achieves 5.36\%, 1.97\%, 11.26\%, 1.81\%, 18.07\%, 8.85\% absolute WER reduction on the DSing\textsuperscript{dev}, DSing\textsuperscript{test}, DALI\textsuperscript{test}, Jamendo, Hansen, and Mauch datasets, compared with the best results among all the previous state-of-the-art approaches respectively. Especially, on DALI\textsuperscript{test}, Hansen, and Mauch datasets, our method significantly exceeds MSTRE-Net \cite{demirel2021mstre} by an average of 12.73\% absolute WER.

\subsection{Effects of pretraining \& finetuning on speech data}

As shown in Fig. \ref{fig1}, wav2vec 2.0 is pretrained and finetuned on speech data before transferring to singing data. The feature representation knowledge learned from speech data is the key to the success of our method. To validate this statement, we conduct ablation studies by comparing the proposed method to two alternative configurations. The first alternative configuration is that we randomly initialize the weights of wav2vec 2.0 and ALT head and then train the whole model on the DSing dataset as per the experiment setup in section 4.2. Note that the first alternative has no transfer learning from the speech domain (without both Stages I and II). The second alternative configuration is that we only perform pretraining on speech data\footnote{https://huggingface.co/facebook/wav2vec2-large-lv60} without Stage II before transferring the wav2vec 2.0 to the singing domain.

\begin{figure}[t]
\begin{center}
\includegraphics[width=\linewidth]{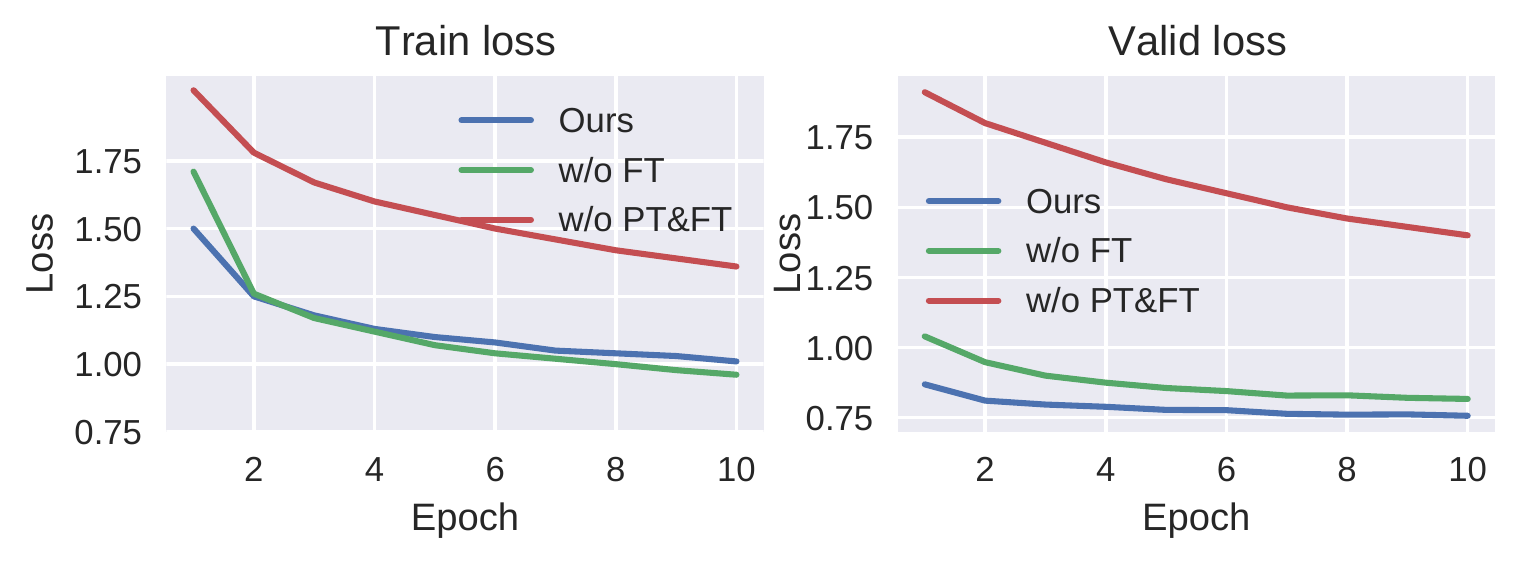}
\caption{Comparison of different training configurations. ``w/o PT \& FT'' refers to without Stages I and II. ``w/o FT'' refers to without Stage II. (Left) Training loss of all configurations for the first 10 epochs; (Right) validation loss of all configurations for the first 10 epochs.}
\label{fig2}
\end{center}
\end{figure}

\begin{figure*}[t]
\begin{center}
\includegraphics[width=1.0\linewidth]{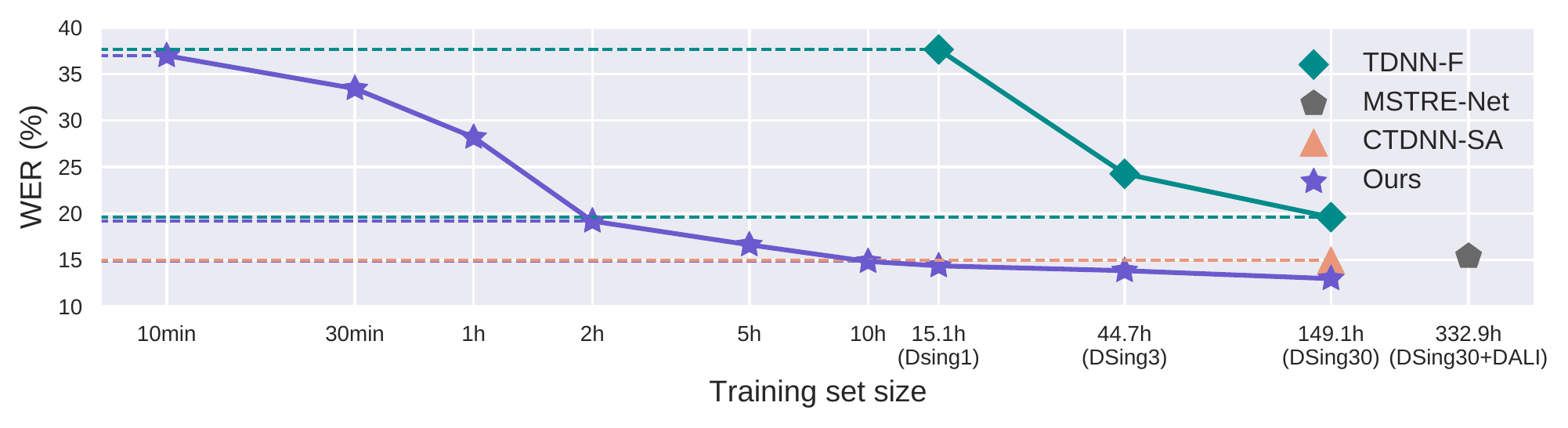}
\vskip -0.2in
\caption{Transcription performance comparison when using different training set sizes, testing on the DSing\textsuperscript{test} dataset.}
\label{fig3}
\end{center}
\end{figure*}

\begin{table}[t]
\centering

\begin{tabular}{@{}l|cc@{}}
\toprule
Method & DSing\textsuperscript{dev} & DSing\textsuperscript{test}\\ 
\midrule
Ours & \textbf{12.34} & \textbf{12.99} \\
\,\, -\,Finetuning & 12.64 ({\color{teal}+\,\,\,0.30}) & 14.58 ({\color{teal}+\,\,\,2.59}) \\
\,\,\,\, -\,Pretraining & 35.61 ({\color{teal}+24.27}) & 39.13 ({\color{teal}+26.14})\\
\bottomrule
\end{tabular}
\caption{WERs (\%) of different training configurations on DSing dataset.}
\label{tab:abl_train}
\end{table}
We evaluate the above training configurations on the DSing dataset. First, we show the curves of training loss and validation loss (loss of the development set) during the training for the first 10 epochs in Fig. \ref{fig2}. The losses are computed through Eq. \ref{eq3}. We observe that without Stages I and II, the training loss and validation loss are much higher than in the other two training configurations. In addition, its convergence is much slower. When we enable Stage I but disable Stage II, the behavior of the training loss is similar to the proposed configuration, except that the training loss is higher at the beginning. However, the validation loss in this setup is higher than that of the proposed configuration. 

We continue training both alternatives until convergence and display the resultant performances in Table \ref{tab:abl_train}. We note that without Stage II, the performance drops by 0.30\% higher WER on DSing\textsuperscript{dev} and 2.59\% higher WER on DSing\textsuperscript{test}. Furthermore, without Stages I and II, the performance degrades severely as WERs on DSing\textsuperscript{dev} and DSing\textsuperscript{test} increase by 24.27\% and 26.14\% respectively, compared to the proposed configuration. The results are consistent with our observations in Fig. \ref{fig2}. Therefore, we conclude that pretraining on speech data plays a significant role in transferring wav2vec 2.0 to the singing domain. Although finetuning on speech data is less crucial than pretraining, it also contributes to empirical performance gains.

\begin{table}[t]
\centering
\begin{tabular}{@{}l|c|cc@{}}
\toprule
Method & DSing\textsuperscript{dev} & DSing\textsuperscript{test}\\ 
\midrule
CTC &  19.86 & 20.99 \\
\,\,+\,S2S &  15.63 ({\color{blue}-4.23}) & 16.95 ({\color{blue}-4.04}) \\
\,\,\,\,+\,LM & \textbf{12.34} ({\color{blue}-7.52)} & \textbf{12.99} ({\color{blue}-8.00)}\\
\bottomrule
\end{tabular}
\caption{WERs (\%) of CTC model and hybrid CTC/attention model on DSing dataset.}
\label{tab:abl_decode}
\end{table}

\subsection{Effects of extending CTC to CTC/attention model}
To validate the effectiveness of changing from CTC to CTC/attention (as in Fig. \ref{fig1}), we compare the performance of our hybrid CTC/attention model with its CTC version, as shown in Table \ref{tab:abl_decode}. We set $\lambda_b=1$ in Eq. \ref{eq5} to disable the branch of the GRU decoder and the S2S linear during the decoding. When $\lambda_c=0$, the RNNLM is disabled.

We observe that the hybrid CTC/attention model achieves better performance by 4.23\% and 4.04\% WER on DSing\textsuperscript{dev} and DSing\textsuperscript{test} respectively than its CTC counterpart, which demonstrates the superiority of our ALT head design. Furthermore, we evaluate the benefits brought by the language model and find that the final model leads to 3.29\% and 3.96\% further absolute WER improvements compared to the hybrid CTC/attention model.

\subsection{Effectiveness of transfer learning in low-resource scenarios}

To explore the effectiveness of our transfer learning method in reducing the amount of required training data, we conduct an ablation study with different training set sizes. We first train our model on DSing30, DSing3, and DSing1, respectively, to observe the performance differences. Then, we further reduce the training set size to a minimum of 10 minutes to create more demanding low-resource setups. We report the WERs achieved on DSing\textsuperscript{test} as the performance measure. When training on 10-minute and 30-minute datasets, the GRU decoder converges too slowly; hence, WERs are computed according to the CTC outputs.

As shown in Fig. \ref{fig3}, our method achieves 14.84\% WER with only 10 hours (about 6.7\% size of DSing30) of labeled singing data, which surpasses the state-of-the-art results of 14.96\% WER achieved by CTDNN-SA \cite{demirel2020automatic} trained on DSing30. It also has better performance with only 2 hours of training data (about 1.3\% size of DSing30) than TDNN-F \cite{dabike2019data:dsing} trained on DSing30 (19.18\% vs. 19.60\% WER). Further, with only 10 minutes of data (1.1\% size of DSing1), our method achieves better results than TDNN-F trained on DSing1 (36.97\% vs. 37.63\% WER). These results demonstrate the feasibility of achieving competitive results with much less training data by adopting the representation knowledge from the speech domain.

\section{Conclusion}
\label{sec:conclusion}

We have introduced a transfer learning approach for the automatic lyric transcription (ALT) task by utilizing the representation knowledge learned by self-supervised learning models on speech data. By performing parameter transfer on wav2vec 2.0 towards the singing domain and extending the original CTC model to a hybrid CTC/attention version, we achieved significant improvement compared to previous state-of-the-art methods on various singing datasets. We demonstrated that both pretraining and finetuning on speech data contribute to the final ALT performance and that pretraining brings more performance gains than finetuning on speech data. Additionally, our method still showed competitive performance using only a tiny proportion of training data, indicating its potential in low-resource scenarios.

\section{Acknowledgement}
\label{sec:acknoledgement}

We would like to thank anonymous reviewers for their valuable suggestions. We also appreciate the help from Emir Demirel, Gabriel Meseguer-Brocal, Jens Kofod Hansen, Chitralekha Gupta for providing datasets. This project is funded in part by a grant (R-252-000-B78-114) from Singapore Ministry of Education.

\bibliography{main}

\begin{thebibliography}{10}
\providecommand{\url}[1]{#1}
\csname url@samestyle\endcsname
\providecommand{\newblock}{\relax}
\providecommand{\bibinfo}[2]{#2}
\providecommand{\BIBentrySTDinterwordspacing}{\spaceskip=0pt\relax}
\providecommand{\BIBentryALTinterwordstretchfactor}{4}
\providecommand{\BIBentryALTinterwordspacing}{\spaceskip=\fontdimen2\font plus
\BIBentryALTinterwordstretchfactor\fontdimen3\font minus
  \fontdimen4\font\relax}
\providecommand{\BIBforeignlanguage}[2]{{%
\expandafter\ifx\csname l@#1\endcsname\relax
\typeout{** WARNING: IEEEtran.bst: No hyphenation pattern has been}%
\typeout{** loaded for the language `#1'. Using the pattern for}%
\typeout{** the default language instead.}%
\else
\language=\csname l@#1\endcsname
\fi
#2}}
\providecommand{\BIBdecl}{\relax}
\BIBdecl

\bibitem{hosoya2005querybysing}
T.~Hosoya, M.~Suzuki, A.~Ito, S.~Makino, L.~A. Smith, D.~Bainbridge, and I.~H.
  Witten, ``Lyrics recognition from a singing voice based on finite state
  automaton for music information retrieval.'' in \emph{ISMIR}, 2005, pp.
  532--535.

\bibitem{fujihara2008audioindexing}
H.~Fujihara, M.~Goto, and J.~Ogata, ``Hyperlinking lyrics: A method for
  creating hyperlinks between phrases in song lyrics.'' in \emph{ISMIR}, 2008,
  pp. 281--286.

\bibitem{gupta2020automatic}
C.~Gupta, E.~Y{\i}lmaz, and H.~Li, ``Automatic lyrics alignment and
  transcription in polyphonic music: Does background music help?'' in
  \emph{ICASSP 2020-2020 IEEE International Conference on Acoustics, Speech and
  Signal Processing (ICASSP)}.\hskip 1em plus 0.5em minus 0.4em\relax IEEE,
  2020, pp. 496--500.

\bibitem{demirel2021low}
E.~Demirel, S.~Ahlb{\"a}ck, and S.~Dixon, ``Low resource audio-to-lyrics
  alignment from polyphonic music recordings,'' in \emph{ICASSP 2021-2021 IEEE
  International Conference on Acoustics, Speech and Signal Processing
  (ICASSP)}.\hskip 1em plus 0.5em minus 0.4em\relax IEEE, 2021, pp. 586--590.

\bibitem{dabike2019data:dsing}
G.~R. Dabike and J.~Barker, ``Automatic lyric transcription from karaoke vocal
  tracks: Resources and a baseline system.'' in \emph{Interspeech}, 2019, pp.
  579--583.

\bibitem{demirel2020automatic}
E.~Demirel, S.~Ahlb{\"a}ck, and S.~Dixon, ``Automatic lyrics transcription
  using dilated convolutional neural networks with self-attention,'' in
  \emph{2020 International Joint Conference on Neural Networks (IJCNN)}.\hskip
  1em plus 0.5em minus 0.4em\relax IEEE, 2020, pp. 1--8.

\bibitem{demirel2021mstre}
E.~Demirel, S.~Ahlb{\"{a}}ck, and S.~Dixon, ``Mstre-net: Multistreaming
  acoustic modeling for automatic lyrics transcription,'' in \emph{Proceedings
  of the 22nd International Society for Music Information Retrieval Conference,
  {ISMIR} 2021, Online, November 7-12, 2021}, J.~H. Lee, A.~Lerch, Z.~Duan,
  J.~Nam, P.~Rao, P.~van Kranenburg, and A.~Srinivasamurthy, Eds., 2021, pp.
  151--158.

\bibitem{gao2022genre}
X.~Gao, C.~Gupta, and H.~Li, ``Genre-conditioned acoustic models for automatic
  lyrics transcription of polyphonic music,'' in \emph{ICASSP 2022-2022 IEEE
  International Conference on Acoustics, Speech and Signal Processing
  (ICASSP)}.\hskip 1em plus 0.5em minus 0.4em\relax IEEE, 2022, pp. 791--795.

\bibitem{sharma2019sing_eval}
B.~Sharma and Y.~Wang, ``Automatic evaluation of song intelligibility using
  singing adapted stoi and vocal-specific features,'' \emph{IEEE/ACM
  Transactions on Audio, Speech, and Language Processing}, vol.~28, pp.
  319--331, 2019.

\bibitem{torrey2009transfer}
L.~Torrey and J.~Shavlik, ``Transfer learning in handbook of research on
  machine learning applications (eds. soria, e., martin, j., magdalena, r.,
  martinez, m. \& serrano, a.) 242--264,'' 2009.

\bibitem{pan2009survey}
S.~J. Pan and Q.~Yang, ``A survey on transfer learning,'' \emph{IEEE
  Transactions on knowledge and data engineering}, vol.~22, no.~10, pp.
  1345--1359, 2009.

\bibitem{sullivan2022improving}
P.~Sullivan, T.~Shibano, and M.~Abdul-Mageed, ``Improving automatic speech
  recognition for non-native english with transfer learning and language model
  decoding,'' \emph{arXiv preprint arXiv:2202.05209}, 2022.

\bibitem{zoph2016transfer}
B.~Zoph, D.~Yuret, J.~May, and K.~Knight, ``Transfer learning for low-resource
  neural machine translation,'' \emph{arXiv preprint arXiv:1604.02201}, 2016.

\bibitem{hsu2021hubert}
W.-N. Hsu, B.~Bolte, Y.-H.~H. Tsai, K.~Lakhotia, R.~Salakhutdinov, and
  A.~Mohamed, ``Hubert: Self-supervised speech representation learning by
  masked prediction of hidden units,'' \emph{IEEE/ACM Transactions on Audio,
  Speech, and Language Processing}, vol.~29, pp. 3451--3460, 2021.

\bibitem{schneider2019wav2vec}
S.~Schneider, A.~Baevski, R.~Collobert, and M.~Auli, ``wav2vec: Unsupervised
  pre-training for speech recognition,'' \emph{arXiv preprint
  arXiv:1904.05862}, 2019.

\bibitem{baevski2020wav2vec}
A.~Baevski, Y.~Zhou, A.~Mohamed, and M.~Auli, ``wav2vec 2.0: A framework for
  self-supervised learning of speech representations,'' \emph{Advances in
  Neural Information Processing Systems}, vol.~33, pp. 12\,449--12\,460, 2020.

\bibitem{riviere2020unsupervised}
M.~Riviere, A.~Joulin, P.-E. Mazar{\'e}, and E.~Dupoux, ``Unsupervised
  pretraining transfers well across languages,'' in \emph{ICASSP 2020-2020 IEEE
  International Conference on Acoustics, Speech and Signal Processing
  (ICASSP)}.\hskip 1em plus 0.5em minus 0.4em\relax IEEE, 2020, pp. 7414--7418.

\bibitem{khurana2022magic}
S.~Khurana, A.~Laurent, and J.~Glass, ``Magic dust for cross-lingual adaptation
  of monolingual wav2vec-2.0,'' in \emph{ICASSP 2022-2022 IEEE International
  Conference on Acoustics, Speech and Signal Processing (ICASSP)}.\hskip 1em
  plus 0.5em minus 0.4em\relax IEEE, 2022, pp. 6647--6651.

\bibitem{meseguer2019data:dali1}
G.~Meseguer-Brocal, A.~Cohen-Hadria, and G.~Peeters, ``Dali: A large dataset of
  synchronized audio, lyrics and notes, automatically created using
  teacher-student machine learning paradigm,'' \emph{arXiv preprint
  arXiv:1906.10606}, 2019.

\bibitem{meseguer2020data:dali2}
------, ``Creating dali, a large dataset of synchronized audio, lyrics, and
  notes,'' \emph{Transactions of the International Society for Music
  Information Retrieval}, vol.~3, no.~1, 2020.

\bibitem{stoller2019end}
D.~Stoller, S.~Durand, and S.~Ewert, ``End-to-end lyrics alignment for
  polyphonic music using an audio-to-character recognition model,'' in
  \emph{ICASSP 2019-2019 IEEE International Conference on Acoustics, Speech and
  Signal Processing (ICASSP)}.\hskip 1em plus 0.5em minus 0.4em\relax IEEE,
  2019, pp. 181--185.

\bibitem{hansen2012recognition}
J.~K. Hansen and I.~Fraunhofer, ``Recognition of phonemes in a-cappella
  recordings using temporal patterns and mel frequency cepstral coefficients,''
  in \emph{9th Sound and Music Computing Conference (SMC)}, 2012, pp. 494--499.

\bibitem{mauch2011integrating}
M.~Mauch, H.~Fujihara, and M.~Goto, ``Integrating additional chord information
  into hmm-based lyrics-to-audio alignment,'' \emph{IEEE Transactions on Audio,
  Speech, and Language Processing}, vol.~20, no.~1, pp. 200--210, 2011.

\bibitem{data:smule}
\BIBentryALTinterwordspacing
Ccrma.Stanford.Edu, ``Smule sing! 300x30x2 dataset,'' Sep. 2018. [Online].
  Available: \url{https://ccrma.stanford.edu/damp/}
\BIBentrySTDinterwordspacing

\bibitem{basak2021end}
S.~Basak, S.~Agarwal, S.~Ganapathy, and N.~Takahashi, ``End-to-end lyrics
  recognition with voice to singing style transfer,'' in \emph{ICASSP 2021-2021
  IEEE International Conference on Acoustics, Speech and Signal Processing
  (ICASSP)}.\hskip 1em plus 0.5em minus 0.4em\relax IEEE, 2021, pp. 266--270.

\bibitem{zhang2021pdaugment}
C.~Zhang, J.~Yu, L.~Chang, X.~Tan, J.~Chen, T.~Qin, and K.~Zhang, ``Pdaugment:
  Data augmentation by pitch and duration adjustments for automatic lyrics
  transcription,'' \emph{arXiv preprint arXiv:2109.07940}, 2021.

\bibitem{gu2022mm}
X.~Gu, L.~Ou, D.~Ong, and Y.~Wang, ``Mm-alt: A multimodal automatic lyric
  transcription system,'' \emph{arXiv preprint arXiv:2207.06127}, 2022.

\bibitem{kruspe2015songified}
A.~M. Kruspe, ``Training phoneme models for singing with "songified" speech
  data,'' in \emph{Proceedings of the 16th International Society for Music
  Information Retrieval Conference, {ISMIR} 2015, M{\'{a}}laga, Spain, October
  26-30, 2015}, M.~M{\"{u}}ller and F.~Wiering, Eds., 2015, pp. 336--342.

\bibitem{pascual2019learning}
S.~Pascual, M.~Ravanelli, J.~Serra, A.~Bonafonte, and Y.~Bengio, ``Learning
  problem-agnostic speech representations from multiple self-supervised
  tasks,'' \emph{arXiv preprint arXiv:1904.03416}, 2019.

\bibitem{chung2019unsupervised}
Y.-A. Chung, W.-N. Hsu, H.~Tang, and J.~Glass, ``An unsupervised autoregressive
  model for speech representation learning,'' \emph{arXiv preprint
  arXiv:1904.03240}, 2019.

\bibitem{jiang2019improving}
D.~Jiang, X.~Lei, W.~Li, N.~Luo, Y.~Hu, W.~Zou, and X.~Li, ``Improving
  transformer-based speech recognition using unsupervised pre-training,''
  \emph{arXiv preprint arXiv:1910.09932}, 2019.

\bibitem{zhang2021transformer}
R.~Zhang, H.~Wu, W.~Li, D.~Jiang, W.~Zou, and X.~Li, ``Transformer based
  unsupervised pre-training for acoustic representation learning,'' in
  \emph{ICASSP 2021-2021 IEEE International Conference on Acoustics, Speech and
  Signal Processing (ICASSP)}.\hskip 1em plus 0.5em minus 0.4em\relax IEEE,
  2021, pp. 6933--6937.

\bibitem{pepino2021emotion}
L.~Pepino, P.~Riera, and L.~Ferrer, ``Emotion recognition from speech using
  wav2vec 2.0 embeddings,'' \emph{arXiv preprint arXiv:2104.03502}, 2021.

\bibitem{zhao2022memobert}
J.~Zhao, R.~Li, Q.~Jin, X.~Wang, and H.~Li, ``Memobert: Pre-training model with
  prompt-based learning for multimodal emotion recognition,'' in \emph{ICASSP
  2022-2022 IEEE International Conference on Acoustics, Speech and Signal
  Processing (ICASSP)}.\hskip 1em plus 0.5em minus 0.4em\relax IEEE, 2022, pp.
  4703--4707.

\bibitem{seo2021wav2kws}
D.~Seo, H.-S. Oh, and Y.~Jung, ``Wav2kws: Transfer learning from speech
  representations for keyword spotting,'' \emph{IEEE Access}, vol.~9, pp.
  80\,682--80\,691, 2021.

\bibitem{fan2020exploring}
Z.~Fan, M.~Li, S.~Zhou, and B.~Xu, ``Exploring wav2vec 2.0 on speaker
  verification and language identification,'' \emph{arXiv preprint
  arXiv:2012.06185}, 2020.

\bibitem{ba2016layer}
J.~L. Ba, J.~R. Kiros, and G.~E. Hinton, ``Layer normalization,'' \emph{arXiv
  preprint arXiv:1607.06450}, 2016.

\bibitem{hendrycks2016gaussian}
D.~Hendrycks and K.~Gimpel, ``Gaussian error linear units (gelus),''
  \emph{arXiv preprint arXiv:1606.08415}, 2016.

\bibitem{vaswani2017attention}
A.~Vaswani, N.~Shazeer, N.~Parmar, J.~Uszkoreit, L.~Jones, A.~N. Gomez,
  {\L}.~Kaiser, and I.~Polosukhin, ``Attention is all you need,''
  \emph{Advances in neural information processing systems}, vol.~30, 2017.

\bibitem{graves2006connectionist}
A.~Graves, S.~Fern{\'a}ndez, F.~Gomez, and J.~Schmidhuber, ``Connectionist
  temporal classification: labelling unsegmented sequence data with recurrent
  neural networks,'' in \emph{Proceedings of the 23rd international conference
  on Machine learning}, 2006, pp. 369--376.

\bibitem{xu2021self}
Q.~Xu, A.~Baevski, T.~Likhomanenko, P.~Tomasello, A.~Conneau, R.~Collobert,
  G.~Synnaeve, and M.~Auli, ``Self-training and pre-training are complementary
  for speech recognition,'' in \emph{ICASSP 2021-2021 IEEE International
  Conference on Acoustics, Speech and Signal Processing (ICASSP)}.\hskip 1em
  plus 0.5em minus 0.4em\relax IEEE, 2021, pp. 3030--3034.

\bibitem{watanabe2017hybrid}
S.~Watanabe, T.~Hori, S.~Kim, J.~R. Hershey, and T.~Hayashi, ``Hybrid
  ctc/attention architecture for end-to-end speech recognition,'' \emph{IEEE
  Journal of Selected Topics in Signal Processing}, vol.~11, no.~8, pp.
  1240--1253, 2017.

\bibitem{chorowski2015attention}
J.~K. Chorowski, D.~Bahdanau, D.~Serdyuk, K.~Cho, and Y.~Bengio,
  ``Attention-based models for speech recognition,'' \emph{Advances in neural
  information processing systems}, vol.~28, 2015.

\bibitem{bahdanau2016end}
D.~Bahdanau, J.~Chorowski, D.~Serdyuk, P.~Brakel, and Y.~Bengio, ``End-to-end
  attention-based large vocabulary speech recognition,'' in \emph{2016 IEEE
  international conference on acoustics, speech and signal processing
  (ICASSP)}.\hskip 1em plus 0.5em minus 0.4em\relax IEEE, 2016, pp. 4945--4949.

\bibitem{defossez2019demucs}
A.~D{\'e}fossez, ``Hybrid spectrogram and waveform source separation,'' in
  \emph{Proceedings of the ISMIR 2021 Workshop on Music Source Separation},
  2021.

\bibitem{ravanelli2021speechbrain}
M.~Ravanelli, T.~Parcollet, P.~Plantinga, A.~Rouhe, S.~Cornell, L.~Lugosch,
  C.~Subakan, N.~Dawalatabad, A.~Heba, J.~Zhong \emph{et~al.}, ``Speechbrain: A
  general-purpose speech toolkit,'' \emph{arXiv preprint arXiv:2106.04624},
  2021.

\bibitem{park2019specaugment}
D.~S. Park, W.~Chan, Y.~Zhang, C.-C. Chiu, B.~Zoph, E.~D. Cubuk, and Q.~V. Le,
  ``Specaugment: A simple data augmentation method for automatic speech
  recognition,'' \emph{arXiv preprint arXiv:1904.08779}, 2019.

\bibitem{kingma2014adam}
D.~P. Kingma and J.~Ba, ``Adam: A method for stochastic optimization,''
  \emph{arXiv preprint arXiv:1412.6980}, 2014.

\end{thebibliography}

\end{document}